\documentstyle[12pt]{article}
\def\singlespace {\smallskipamount=3.75pt plus1pt minus1pt
                  \medskipamount=7.5pt plus2pt minus2pt
                  \bigskipamount=15pt plus4pt minus4pt
                  \normalbaselineskip=15pt plus0pt minus0pt
                  \normallineskip=1pt
                  \normallineskiplimit=0pt
                  \jot=3.75pt
                  {\def\smallskip {\vskip\smallskipamount}}
                  {\def\medskip   {\vskip\medskipamount}}
                  {\def\bigskip   {\vskip\bigskipamount}}
                  {\setbox\strutbox=\hbox{\vrule 
                    height10.5pt depth4.5pt width 0pt}}
                  \parskip 7.5pt
                  \normalbaselines}
\def\middlespace {\smallskipamount=5.625pt plus1.5pt minus1.5pt
                  \medskipamount=11.25pt plus3pt minus3pt
                  \bigskipamount=22.5pt plus6pt minus6pt
                  \normalbaselineskip=22.5pt plus0pt minus0pt
                  \normallineskip=1pt
                  \normallineskiplimit=0pt
                  \jot=5.625pt
                  {\def\smallskip {\vskip\smallskipamount}}
                  {\def\medskip   {\vskip\medskipamount}}
                  {\def\bigskip   {\vskip\bigskipamount}}
                  {\setbox\strutbox=\hbox{\vrule 
                    height15.75pt depth6.75pt width 0pt}}
                  \parskip 11.25pt
                  \normalbaselines}
\def\doublespace {\smallskipamount=7.5pt plus2pt minus2pt
                  \medskipamount=15pt plus4pt minus4pt
                  \bigskipamount=30pt plus8pt minus8pt
                  \normalbaselineskip=30pt plus0pt minus0pt
                  \normallineskip=2pt
                  \normallineskiplimit=0pt
                  \jot=7.5pt
                  {\def\smallskip {\vskip\smallskipamount}}
                  {\def\medskip   {\vskip\medskipamount}}
                  {\def\bigskip   {\vskip\bigskipamount}}
                  {\setbox\strutbox=\hbox{\vrule 
                    height21.0pt depth9.0pt width 0pt}}
                  \parskip 15.0pt
                  \normalbaselines}

\begin{document}

\author{T. P. Singh  \\ 
Tata Institute of Fundamental Research\\Homi Bhabha Road\\Mumbai 400 005,
INDIA\\e-mail: tpsingh@tifrc3.tifr.res.in}
\title{Gamma-Ray Bursts and \\ Quantum Cosmic 
Censorship\footnote{This essay received the third award from the
Gravity Research Foundation for the year 1998}}
\date{}
\maketitle

\begin{abstract}
\noindent Gamma-ray bursts are believed to result from the coalescence of
binary neutron stars. However, the standard proposals for conversion of the
gravitational energy to thermal energy have difficulties. We show that if
the merger of the two neutron stars results in a naked singularity, instead
of a black hole, the ensuing quantum particle creation can provide the
requisite thermal energy in a straightforward way. The back-reaction of the
created particles can avoid the formation of the naked singularity predicted
by the classical theory. Hence cosmic censorship holds in the quantum
theory, even if it were to be violated in classical general relativity.
\end{abstract}

\newpage\ \middlespace

\noindent Gamma-ray bursts (GRBs) are non-thermal bursts of low energy $%
\gamma $-rays. The detection of isotropy in their distribution by the BATSE
detector on COMPTON-GRO, together with the recent detection of afterglows
following some bursts, strongly supports their cosmological origin. A burst
takes place about once every million years in a galaxy, and an energy of
about $10^{51}-10^{52}$ ergs is released in every burst, making GRBs the
most luminous objects in the Universe. Causality requirements restrict the
size of the initial source that triggers a burst to about $10^7$ cm. The
time profile of a burst is very intricate, and there is great diversity in
the temporal structure of bursts. It is hence a fascinating challenge for
theoretical astrophysicists to construct a model which can correctly explain
these observations.

The deposition of such a large amount of energy by the progenitor inside a
small volume gives rise to photon energy densities which result in an
optically thick, thermal, $\gamma e^{\pm }$ fireball. Such a compact energy
deposition results in a highly relativistic expansion, and a $\Gamma $
factor of $100-1000$ can be achieved. This is the standard fireball model,
which to a good degree is independent of the nature of the progenitor. The
expanding fireball becomes optically thin at a later stage, and the kinetic
energy of expansion is converted into the observed non-thermal $\gamma $-ray
photons by a dissipative mechanism like shocks. These could be external
shocks, resulting from the interaction of the outflow with an external
medium like the ISM, or internal shocks which take place in the outflow
itself. It has recently been argued \cite{sari} that the external shock
model is not viable, as these shocks cannot produce the complex, irregular
temporal structure. The shocks, if assumed to be internal, carry information
about the progenitor of the burst.

The most attractive model for the progenitor is the coalescence of binary
neutron stars, which can release an energy of the order of $10^{53}$ ergs.
Observations of pulsars suggest that such neutron star mergers take place
about once every million years in a galaxy, which is in remarkable agreement
with the observed GRB rate. However, a key problem with the neutron star
merger model is: how is the gravitational energy available from the merger
converted into a fireball? Two solutions have been proposed. The first is
that some of the energy released as neutrinos is reconverted, by the
collision of these neutrinos, into $e^{+}e^{-}$ pairs, or into photons.
However, simulations suggest that this process is inefficient \cite{ruf} and
hence unlikely. The second solution is that strong magnetic fields ($\sim
10^{15}$ Gauss) convert the rotational energy of the system into a
relativistic outflow. Though this solution cannot be ruled out, it remains
an open issue as to how such large magnetic fields could be generated.

While considering the above two options for energy generation, one tacitly
assumes that the merger of two neutron stars eventually leads to the
formation of a black hole. The purpose of the present essay is to propose an
alternative mechanism for energy generation from the merger, by questioning
the assumption that the merger results in a black hole. It is customary in
astrophysics to assume that continual gravitational collapse results in
black hole formation. However, it is known to general relativists that this
depends very crucially on the cosmic censorship hypothesis being valid. The
hypothesis states that generic singularities resulting from gravitational
collapse are hidden behind horizons and are not naked. As we explain below,
a naked singularity has observational properties that are extremely
different from those of a black hole.

In this essay, we assume that the coalescence of the two neutron stars
results in the formation of a naked singularity, and not a black hole. We
then show that such an assumption can successfully solve the energy
generation problem in a natural manner. Striking evidence that the merger
might actually result in a naked singularity comes from a recent work \cite
{baum}, where it is found that the ratio of the angular momentum to mass can
be greater than one for the collapsed object. As is well known, in this
range the collapse results in a naked singularity, and not a Kerr black
hole. While we cannot rule out the possibility that the system can still
lose enough angular momentum (say via gravitational radiation) before the
completion of collapse, we find it quite attractive to consider the
implications for a GRB if a naked singularity were indeed to form.

Before considering these implications we would like to make contact with
what is known about naked singularities in classical general relativity.
Till today, the cosmic censorship hypothesis remains unproven. However, a
few examples of naked singularity formation have been found over the last
few years, in general relativistic studies of gravitational collapse. Most
of these examples have been found in studies of spherical collapse using
equations of state corresponding to dust, perfect and imperfect fluids \cite
{sin}. It is typically found that for a given equation of state, both
black-hole and naked singularity solutions result, depending on the choice
of initial data. But perhaps the most striking evidence for censorship
violation comes from studies of spherical collapse of a massless scalar
field \cite{chop}. It has been found that regions of arbitrarily high
curvature resulting in collapse are visible from infinity. While the actual
naked singularity in this model is non-generic, regions of unbounded
curvature (i.e. the black holes of arbitrarily small mass) result from
generic initial data. The small mass black holes violate the spirit of
cosmic censorship, because one can see as close to the singularity as one
desires.

It is only fair to say at this stage that the censorship hypothesis has not
been disproved, because the examples studied so far involve one or the other
special form of matter. Nonetheless, the very fact that some examples have
been discovered is sufficient cause for enquiring if there are any known
astrophysical phenomena which can be modelled after naked singularities.
Moreover, even if the censorship hypothesis were to be disproved some day,
naked singularities will remain abstract curiosities unless some
observational evidence is found in their favour. After all, black holes
themselves became acceptable only after excellent evidence in their favour
was found in systems like X-ray binaries and the centres of galaxies.

What will a naked singularity look like to an observer who is watching the
collapse from far away? There is good theoretical evidence that even when a
singularity forming in collapse is naked, the outgoing light rays starting
from the singularity are infinitely redshifted. Similarly, even if regions
of unbounded high curvature are in principle visible, like in Choptuik's
study, light leaving these regions will be extremely redshifted. Naked
singularities are thus effectively black, making them indistinguishable from
black holes, so long as only classical processes are considered.

However, quantum effects will be of fundamental importance prior to the
onset of a naked singularity. This is so even before the quantum
gravitational regime is approached, because the development of high
spacetime curvatures will give rise to intense quantum particle creation.
The nature of this particle creation in a naked singular spacetime is
fundamentally different from the Hawking radiation that accompanies a
quantum black hole, and this difference serves to distinguish a naked
singularity from a black hole, observationally. We explain this using an
example developed in \cite{eard}, where quantum effects were studied in the
naked singular spacetime resulting from spherical null dust collapse. It is
found that the outgoing flux of quantum radiation diverges on the Cauchy
horizon, as seen by a far away observer. (This is in spite of the infinite
classical redshift referred to above). Most of the quantum particle creation
takes place during a very short period prior to the formation of the Cauchy
horizon. Similar features have been observed in other studies \cite{cva} 
\cite{other}, and these features are expected to be generic to naked
singularities.

Thus, a quantum naked singularity is a burst like phenomenon, unlike the
black hole which evaporates slowly because of the Hawking radiation.
Further, the divergence of the stress tensor on the Cauchy horizon
represents a quantum blue shift instability, and strongly suggests that
back-reaction will avoid formation of the naked singularity. Hence, even if
censorship is violated in classical general relativity, it will hold in
quantum theory - this is the quantum cosmic censorship \cite{cva} alluded to
in the title.

We see that a quantum naked singularity bears resemblance to a gamma-ray
burst. If the neutron star merger results in a naked singularity, quantum
particle creation can convert a good fraction of the original infalling
material into outgoing radiation. In fact, for every solar mass of matter,
the equivalent energy is $M_{sun}c^2\approx 10^{54}$ ergs, which will be
available via quantum pair production. Besides, photons and neutrinos will
be produced with equal probability, hence the required energy deposit will
be available as electromagnetic radiation. This offers a natural solution to
the energy conversion problem. While a good deal of work will have to be
done to test the naked singularity - GRB model, it is an avenue that appears
worth pursuing further (see also \cite{wit}).

The coalescing binary neutron star system is an excellent testing laboratory
for the cosmic censorship hypothesis. If the merger results in a naked
singularity, the gravitational wave signal will be significantly different
from what would be seen by LIGO and VIRGO if a black hole forms. The gravity
wave emission from a naked singularity will be much more copious and long
lasting, because now regions of very high curvature will participate in
producing the waves. If gamma-ray bursts indeed result from the formation of
a naked singularity at the end of the merger, they are also ideal testing
grounds for the quantum gravitational effects that become important in the
approach to a naked singularity. These bursts are perhaps the only system
nature has offered us in which candidate quantum gravity theories like
string theory and quantum general relativity can be tested experimentally.

I acknowledge partial support of the {\it Junta Nacional de Investigac\~ao
Cient\'ifica e Tecnol\'ogica} (JNICT) Portugal, under contract number
CERN/S/FAE/1172/97.

\end{document}